\documentclass[twocolumn,aps,superscriptaddress,nofootinbib,floatfix]{revtex4-2}

\usepackage{amsfonts}
\usepackage{amsmath}
\usepackage{float}
\usepackage{comment}
\usepackage{placeins}
\usepackage{graphicx}
\usepackage[hidelinks]{hyperref}
\usepackage{color,amsmath,amssymb,bm}
\usepackage[sort,compress]{cleveref}

\crefname{section}{Sec.\!}{Secs.\!}
\crefname{equation}{Eq.\!}{Eqs.\!}
\crefname{figure}{Fig.\!}{Figs.\!}
\crefname{table}{Tab.\!}{Tabs.\!}
\crefname{appendix}{App.\!}{Apps.\!}
\crefname{chapter}{Chapter}{Chapters}
\usepackage{tensor}
\usepackage{lipsum}
\def\d{\,\mathrm{d}}

\usepackage[normalem]{ulem}  

\renewcommand{\sout}{\bgroup \color{black} \ULdepth=-.5ex \ULset}

\DeclareMathOperator{\Tr}{Tr}

\def\blfootnote{\xdef\@thefnmark{}\@footnotetext}

\newcommand{\beq}{\begin{equation}}
\newcommand{\eeq}{\end{equation}}
\newcommand{\bea}{\begin{eqnarray}}
\newcommand{\eea}{\end{eqnarray}}

\begin{document}

\title{Elliptic flow of charm quarks produced in the early stage of pA collisions}

\author{Gabriele Parisi}
\email{gabriele.parisi@dfa.unict.it}
\affiliation{Department of Physics and Astronomy, University of Catania, Via Santa Sofia 64, 1-95125 Catania, Italy}
\affiliation{Laboratori Nazionali del Sud, INFN-LNS, Via Santa Sofia 62, I-95123 Catania, Italy}

\author{Fabrizio Murgana}
\email{fabrizio.murgana@dfa.unict.it}
\affiliation{SUBATECH, Université de Nantes, IMT Atlantique, IN2P3/CNRS, 4 rue Alfred Kastler, 44307 Nantes cedex 3, France}

\author{Vincenzo Greco}
\email{greco@lns.infn.it}
\affiliation{Department of Physics and Astronomy, University of Catania, Via Santa Sofia 64, 1-95125 Catania, Italy}
\affiliation{Laboratori Nazionali del Sud, INFN-LNS, Via Santa Sofia 62, I-95123 Catania, Italy}

\author{Marco Ruggieri}
\email{marco.ruggieri@dfa.unict.it}
\affiliation{Department of Physics and Astronomy, University of Catania, Via Santa Sofia 64, 1-95125 Catania, Italy}
\affiliation{INFN-Sezione di Catania, Via Santa Sofia 64, I-95123 Catania, Italy}

\date{\today}

\begin{abstract}
We investigate the build-up of elliptic flow of charm quarks produced in the early pre-equilibrium stage of high-energy proton--nucleus collisions. The initial stage is modeled within the Color Glass Condensate framework as an evolving glasma, initialized through the McLerran--Venugopalan model. Subnucleonic fluctuations have been implemented as constituent-quark hotspots for both the proton and the nuclear participants. Charm quarks are propagated in the evolving non-Abelian background by solving the relativistic Wong equations for their coordinates, momenta, and color charges.
First, we compute the nuclear modification factor of 
charm quarks, finding a slight migration 
towards higher $p_T$ states in agreement
with previous results in the literature. 
Then, we focus on the azimuthal anisotropies acquired through the interaction with glasma fields. We find that glasma-induced momentum anisotropies are efficiently transmitted to heavy quarks within $\tau \sim 0.4~\mathrm{fm/c}$, leading to a sizeable charm-quark $v_2$, 
with a magnitude that increases with the strength of the initial fields and with the number of nuclear participants.
Remarkably, we show that the early-stage contribution alone can account for a significant fraction of the experimentally observed $J/\psi$ elliptic flow in p-Pb collisions, indicating that pre-hydrodynamic dynamics can play a non-negligible role in the final-state heavy-flavor collectivity, especially in small systems.
\end{abstract}

\maketitle

\section{Introduction}
\label{sec:introduction}

Ultra-relativistic Heavy-Ion Collisions (HICs) provide a unique opportunity to investigate Quantum ChromoDynamics (QCD) under extreme conditions of energy density and strong color fields. A key outcome of the last decades of experimental and theoretical progress is the emerging picture that the earliest stage of the collision is dominated by highly occupied gluonic degrees of freedom: these form a medium called glasma, its structure and real-time evolution can be effectively described within the Color Glass Condensate (CGC) framework \cite{McLerran:1993ni,McLerran:1993ka,McLerran:1994vd, Lappi:2006fp}, see \cite{Gelis:2010nm,Iancu:2003xm,McLerran:2008es,Gelis:2012ri} for reviews. In particular, during the first $\sim 0.4~\mathrm{fm/c}$ after the impact, the system is expected to be far from equilibrium and characterized by strong, coherent chromo-electromagnetic fields, whose dynamics controls the initial momentum-space anisotropies and sets the stage for the later emergence of collective phenomena.

In this context, Heavy Quarks (HQ) constitute especially valuable probes of the early-time dynamics \cite{Dong:2019unq}: being produced predominantly in initial hard scatterings, their formation time is short and their production rates can be described within perturbative QCD (pQCD). As a consequence, heavy quarks experience the glasma fields from the very early stages of the collision and retain sensitivity to the pre-equilibrium evolution, offering a clean handle to quantify the amount of anisotropy and collectivity generated prior to the onset of hydrodynamization.

In this work, we compute the elliptic flow of charm
quarks produced in the early stage of proton-proton and proton-nucleus collisions
at the LHC energy.
The early stage is modeled as an evolving glasma, namely,
as a set of intense and correlated gluon fields,
whose dynamics is governed by the classical Yang-Mills equations. 
Charm quarks are then coupled as probes to the
evolving bulk fields, and their dynamics is studied
by solving the relevant relativistic kinetic theory equations,
namely the Wong equations~\cite{Wong:1970fu, Heinz:1984yq}.
As an improvement of previous works on the subject \cite{Oliva:2024rex, Parisi:2025slf, Bazak:2023kol, Carrington:2020ssh, Carrington:2021qvi, Avramescu:2023qvv, Avramescu:2024poa, Avramescu:2024xts}, we introduce nucleonic fluctuations on both the proton and
the nucleus side, while in most of the previous works
a homogeneous probability distribution of color charges was used
on both colliding projectiles.
Previous works investigated the
elliptic flow of the bulk gluon fields using inhomogeneous
color charge distributions \cite{Schenke:2015aqa, Carrington:2024utf}, 
but these focused on the bulk anisotropies in AA and pA collisions,
while here we focus also on how these anisotropies are transmitted to charm quarks.

A finite momentum anisotropy is known to emerge already at early times~\cite{Schenke:2015aqa} and is 
not related to a hydro-like flow of the bulk; rather, it is
intrinsic to the domain-like structure of the glasma in the transverse plane. 
Indeed, the classical gauge fields develop correlation domains characterized by a finite transverse correlation length, such that within each domain one can identify, on average, a preferred direction
of the color fields, 
thereby locally breaking rotational symmetry.

One of our main results is 
that the
momentum anisotropies of the bulk gluon fields
are efficiently transmitted to heavy quarks within $\tau \sim 0.4~\mathrm{fm/c}$, leading to a sizeable charm-quark $v_2$.
We also show that the early-stage contribution alone can 
account for a significant fraction of the experimentally observed $J/\psi$ elliptic flow in p-Pb collisions, 
indicating that pre-hydrodynamic dynamics can play a non-negligible role in the final-state heavy-flavor collectivity.

The plan of the article is the following.
In Section II we review the glasma picture and
its evolution on a real-time lattice within a
gauge-invariant formulation.
In Section III we describe the framework used to couple
the charm quarks to the bulk fields in the early stage.
In Section IV we briefly show results for the nuclear
modification factor of charm quarks.
In Section V we analyze the elliptic flow that the bulk glasma fields and the charm quarks evolving therein accumulate during the early stage.
Finally, in Section VI we draw our conclusions.

\section{The evolving glasma}
\label{The evolving Glasma}

Several different strategies have been used to understand and describe the earliest phase of relativistic heavy-ion collisions. One which is commonly applied is the Color Glass Condensate effective theory \cite{McLerran:1993ni,McLerran:1993ka,McLerran:1994vd}, which is based on a separation of scales between the high Bjorken-$x$ degrees of freedom, i.e. the valence partons, and the low $x$ degrees of freedom which they generate, which are soft gluons.

\subsection{Color charges for p and A}
\label{subsec:Color_charges_for_p_and_A}
In our work, to realize the aforementioned separation,
the sources of the initial gluon fields 
are generated using the McLerran-Venugopalan (MV) model, 
in which the large-$x$ color sources 
are randomly distributed
on an infinitely thin color sheet.
For each point in transverse space, the event-by-event distribution of these charges is gaussian, characterized by
zero average 
\begin{equation}
    \langle \rho^a (\mathbf{x}_\perp)\rangle=0,
    \label{1}
\end{equation}
and variance given by
\begin{equation}
    \langle \rho^a (\mathbf{x}_\perp)\rho^b (\mathbf{y}_\perp)\rangle=g^2 \mu^2 \delta^{ab} \delta^{(2)}(\mathbf{x}_\perp-\mathbf{y}_\perp).
    \label{2}
\end{equation}
Here, $g$ is the coupling constant (we use $g=2$,
corresponding to $\alpha_s=0.3$), 
and $\mu$ is the so-called MV parameter, describing the number density of the color charges per unit
of area in the transverse plane.
In our implementation, we relax the single-sheet hypothesis of the 
original MV model 
by generating a certain number $N_s$ of color sheets stacked on top of one another \cite{LAPPI:2010ykh}. 
Consequently, for each of the sheets,
the numerical implementation of Eq.~\eqref{2}
we will use in this work is
\begin{equation}
    \langle \rho^a_{n,x}\rho^b_{m,y}\rangle=g^2 \mu^2 \frac{\delta_{n,m}}{N_s}\delta^{a,b} \frac{\delta_{x,y}}{a_\perp^2},
    \label{3}
\end{equation}
where $a,b=1,\dots, N_c^2-1$ are $SU(3)$ color indices, $n,m=1,\dots,N_s$ are the color sheet indices and $x,y$ span each point of a $N_\perp\times N_\perp$ lattice, whose transverse length is $L_\perp$ and lattice spacing is $a_\perp=L_\perp/N_\perp$. Operatively, the conditions \eqref{1} and \eqref{3} are satisfied by generating random Gaussian numbers with mean zero and standard deviation equal to $\sqrt{(g^2\mu^2)/(N_s a_\perp^2)}$.
Below we explain how to build up the gluon fields generated
by each sheet, and how to
combine all these fields to obtain the initial
gluon field in the glasma.
Unless differently stated, we will use $N_s=50$ in Eq.~\eqref{3}
throughout this work: we have seen that the observed dependence of the energy density on the number of color sheets lasts until this value of $N_s$. Moreover, such a value smears the statistical fluctuations due to the aforementioned random charge extraction.\\

These steps are shared by the charge generation of both a proton and a heavy ion. Indeed, the original MV model assumed a source charge density that was homogeneous in the transverse plane, which is feasible when dealing with large-A nuclei, of which we can neglect the details at the level of single nucleons. On the other hand, a more reasonable description of protons and nuclei involves a varying nuclear density in the transverse plane, in order to take into account for the underlying quark structure. The main difference between those two cases lies in the choice of $\mu$. In what we will call the ‘wall' case, $\mu$ is chosen as a constant, which translates into uniform fluctuations of color charge throughout the lattice. The proton case (which we will soon generalize for all nuclei) is instead more involved. Let us introduce the thickness function of the proton, $T_p$, as
\begin{equation}
T_p(\mathbf{x}_\perp) = 
\frac{1}{3} \sum_{i=1}^3 \frac{1}{2\pi B_q} \exp\left[-\frac{(\mathbf{x}_\perp-\mathbf{x}_i)^2}{2B_q}\right],
\label{eq:bd1}
\end{equation}
where the $\mathbf{x}_i$ denote the positions of the constituent quarks, which are randomly extracted from the distribution
\begin{equation}
T_{cq}(\mathbf{x}_\perp) = \frac{1}{2\pi B_{cq}}
\exp\left(-\frac{\mathbf{x}_\perp^2}{2B_{cq}}\right).
\label{eq:bd2}
\end{equation}
Parameters have been chosen as $B_q=0.3$ GeV$^{-2}$, $B_{cq}=4$ GeV$^{-2}$, i.e. the widths of the two gaussians differ by a factor around 4. After doing so, we evaluate the saturation scale $Q_s$ in this model as
\begin{equation}
Q_s^2(x,\mathbf{x}_\perp) = \frac{2\pi^2}{N_c}  \alpha_s\cdot 
xg(x,Q_0^2)\cdot T_p(\mathbf{x}_\perp),
\label{eq:qs_2_77}
\end{equation}
from which $\mu(x,\mathbf{x}_\perp)$ of the proton is given by \cite{Schenke:2020mbo}
\begin{equation}
g^2\mu(x,\mathbf{x}_\perp) = c Q_s(x,\mathbf{x}_\perp),
\label{eq:gmu_ajk}
\end{equation}
with $c=1.25$. Note that by virtue of Eq.~\eqref{eq:qs_2_77} not only $\mu$, but also $Q_s$ depend on the transverse plane coordinates.
Also, in principle $xg$ in \eqref{eq:qs_2_77} should be computed at the scale $Q_s$ by means of the DGLAP equation with a proper initialization. This was done in \cite{Schenke:2020mbo,Rezaeian:2012ji}, and we reserve this approach for future studies. For the sake of simplicity, in the present study we limit ourselves by assuming that $xg$ is given by the initial condition at 
$Q_0^2=1.51$ GeV$^2$ as in \cite{Schenke:2020mbo,Rezaeian:2012ji}, namely
\begin{equation}
xg(x,Q_0^2)=A_g x^{-\lambda_g}(1-x)^{f_g},
\label{eq:anna23}
\end{equation}
with $A_g=2.308$, $\lambda_g=0.058$ and $f_g=5.6$. This simplification is partly justified by the fact that the average saturation scale for the proton is of the order of $Q_s\sim 1$ GeV, hence we do not expect the DGLAP evolution of the $xg$ in Eq.~\eqref{eq:anna23} to lead to significant changes. For pA collisions at the LHC energy, the relevant values of $x$ are in the range $[10^{-4},10^{-3}]$. In our calculations, we will thus fix $x$ in the aforementioned range, then compute $xg$ by virtue of Eq.~\eqref{eq:anna23}. If not otherwise stated, we will use the corresponding value obtained for $x=10^{-4}$, that is $xg=3.94$.\\

For nuclei larger than a proton, we can repeat the above procedure 
for $N_\mathrm{part}$ nucleons, namely, for each nucleon we firstly extract
the center of the three hotspots by the distribution~\eqref{eq:bd2};
then we extract the color charges
that spread around the three hotspots
by virtue of Eq.~\eqref{eq:bd1}.
The total number of hotspots that need to be extracted is thus
$N_\mathrm{HS} = 3N_\mathrm{part}$.
The thickness function for the nucleus side can thus
be written as
\begin{equation}
T_p(\mathbf{x}_\perp) = 
\frac{N_\mathrm{part}}{N_\mathrm{HS}} \sum_{i=1}^{N_\mathrm{HS}} \frac{1}{2\pi B_q} \exp\left[-\frac{(\mathbf{x}_\perp-\mathbf{x}_i)^2}{2B_q}\right],
\label{eq:bd1nucleus}
\end{equation}
which is correctly normalized as
\begin{equation}
\int d^2\mathbf{x}_\perp T_p(\mathbf{x}_\perp) = N_\mathrm{part},
\label{eq:normalizationTPp}
\end{equation}
and reproduces Eq.~\eqref{eq:bd1} in the limit $N_\mathrm{part}=1$.
In our calculations for pA, changing $N_\mathrm{part}$ amounts to
change the number of participants from the nucleus, 
that is, the centrality of the collision. 
Once the $T_p$ is computed, the 
$Q_s^2$ and the $\mu$ are
obtained via Eqs.~\eqref{eq:qs_2_77}
and~\eqref{eq:gmu_ajk}. 

\begin{figure*}[t!]
    \centering
\includegraphics[width=0.31\linewidth]{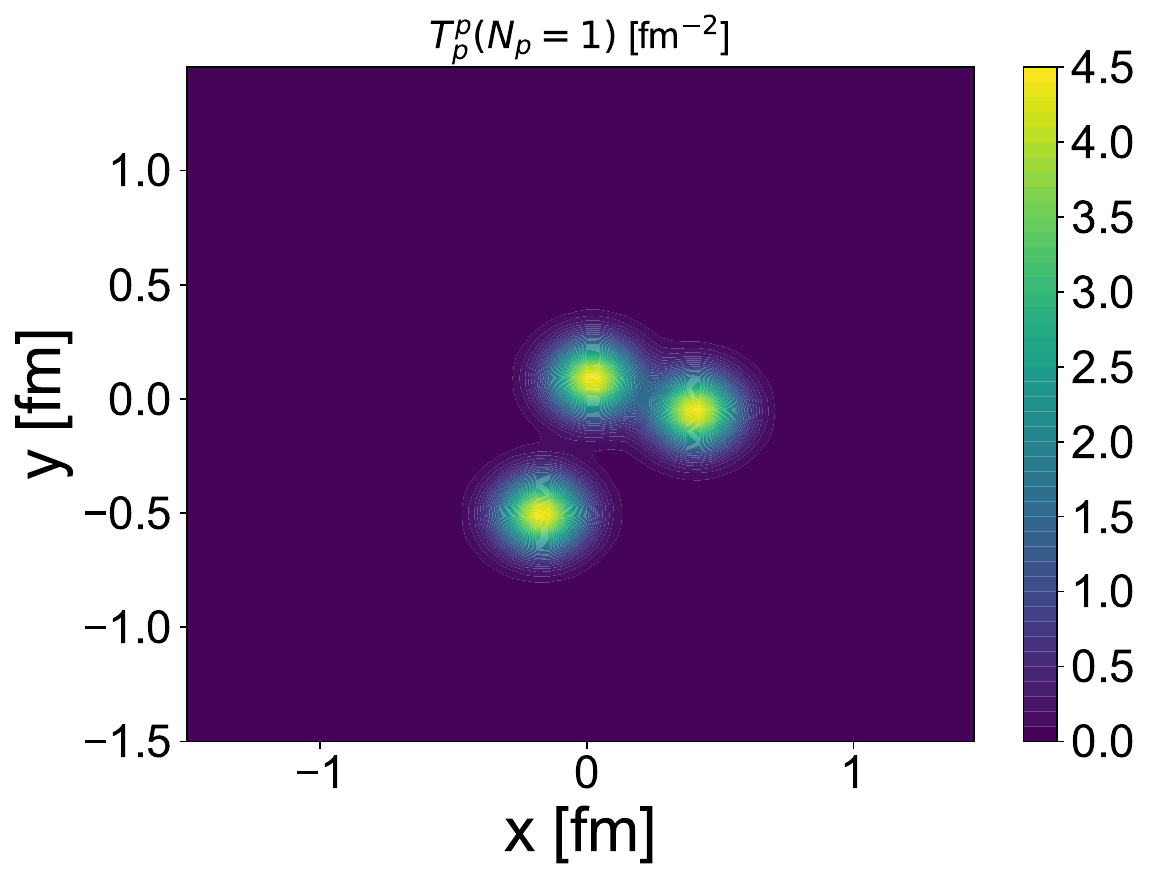}~~
\includegraphics[width=0.32\linewidth]{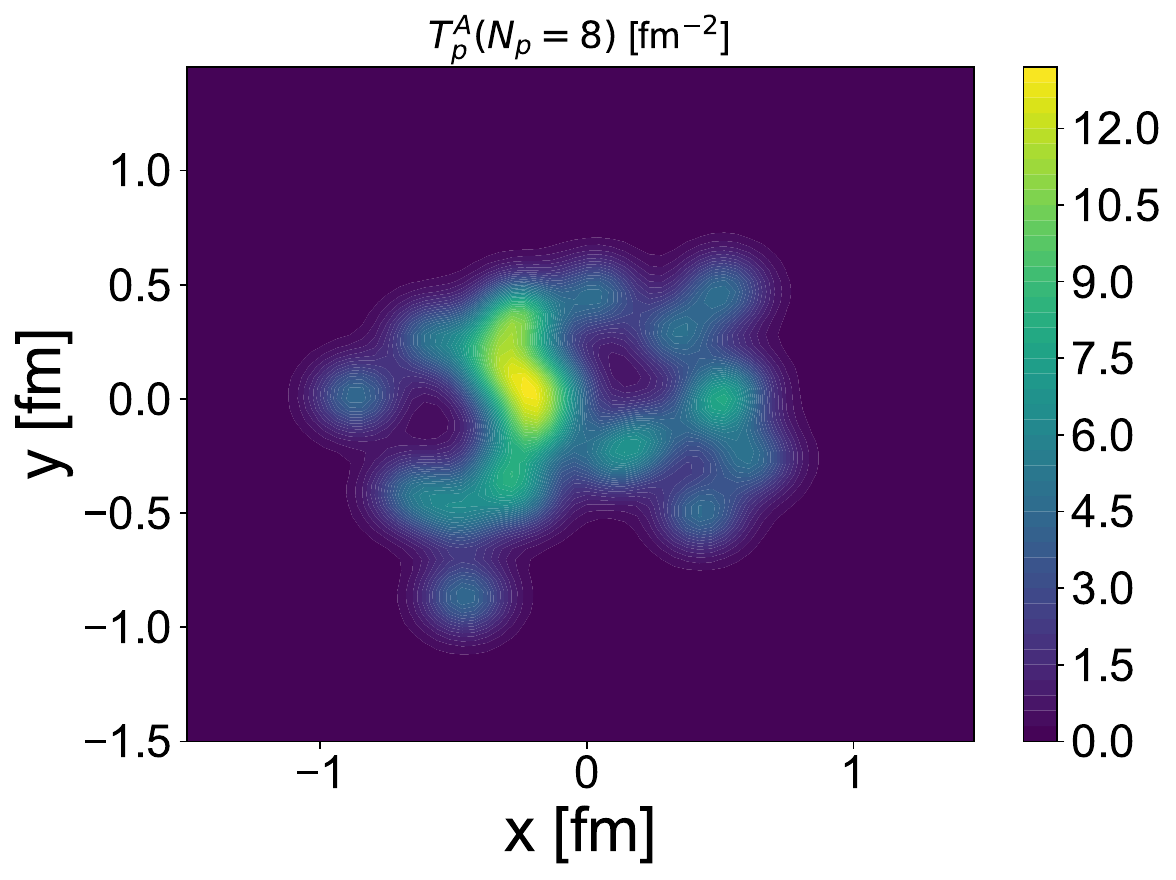}~~
\includegraphics[width=0.31\linewidth]{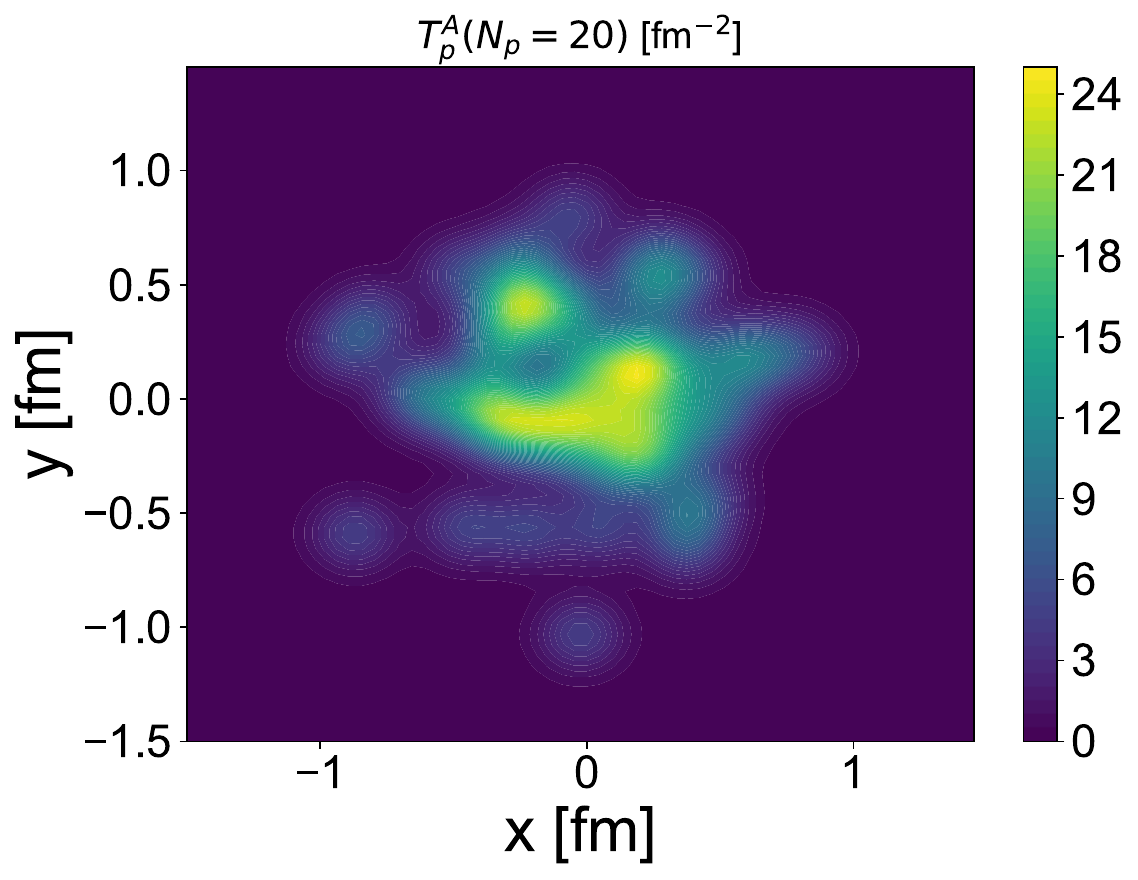}\\
\includegraphics[width=0.3\linewidth]{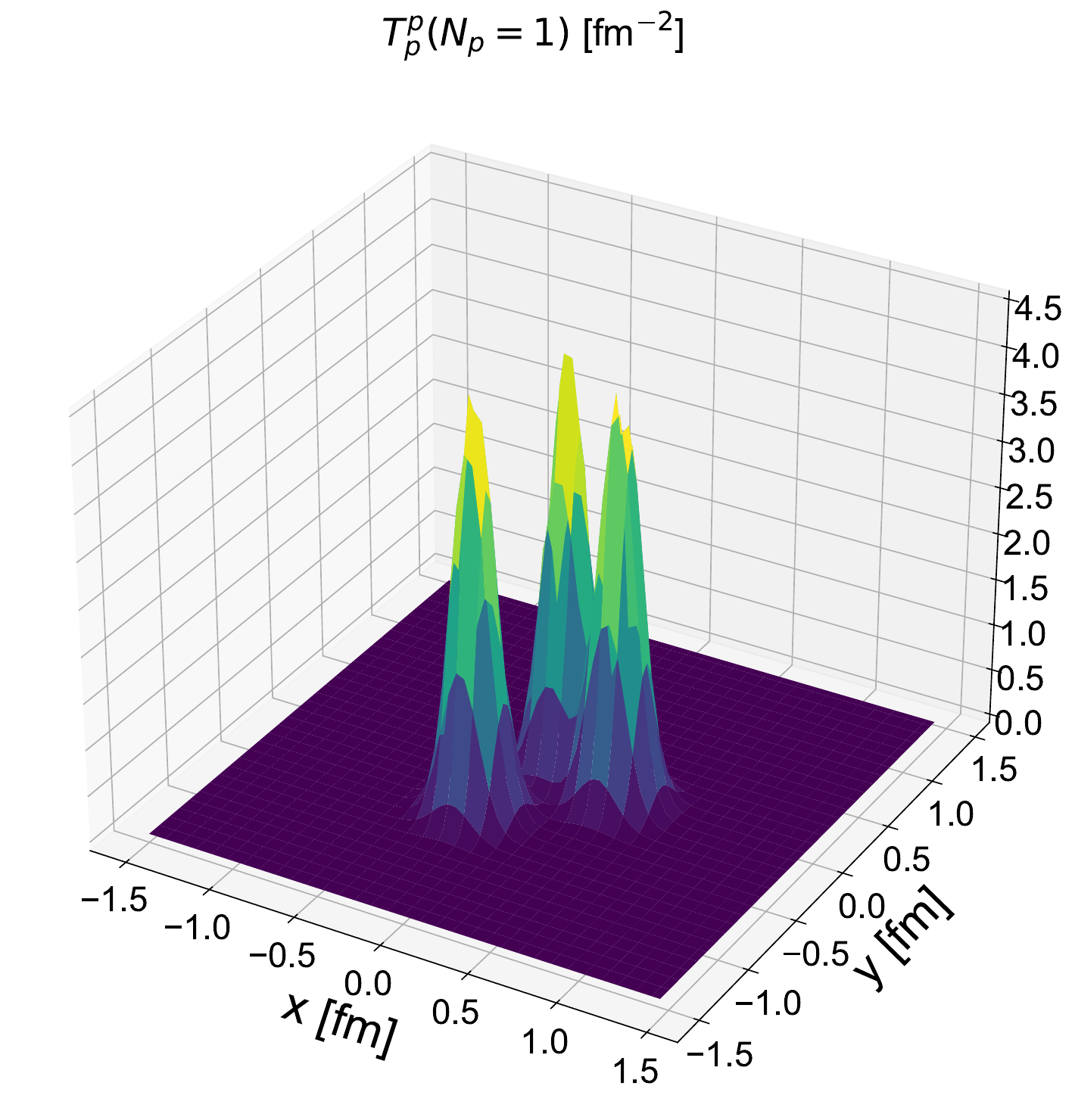}~~~~
\includegraphics[width=0.3\linewidth]{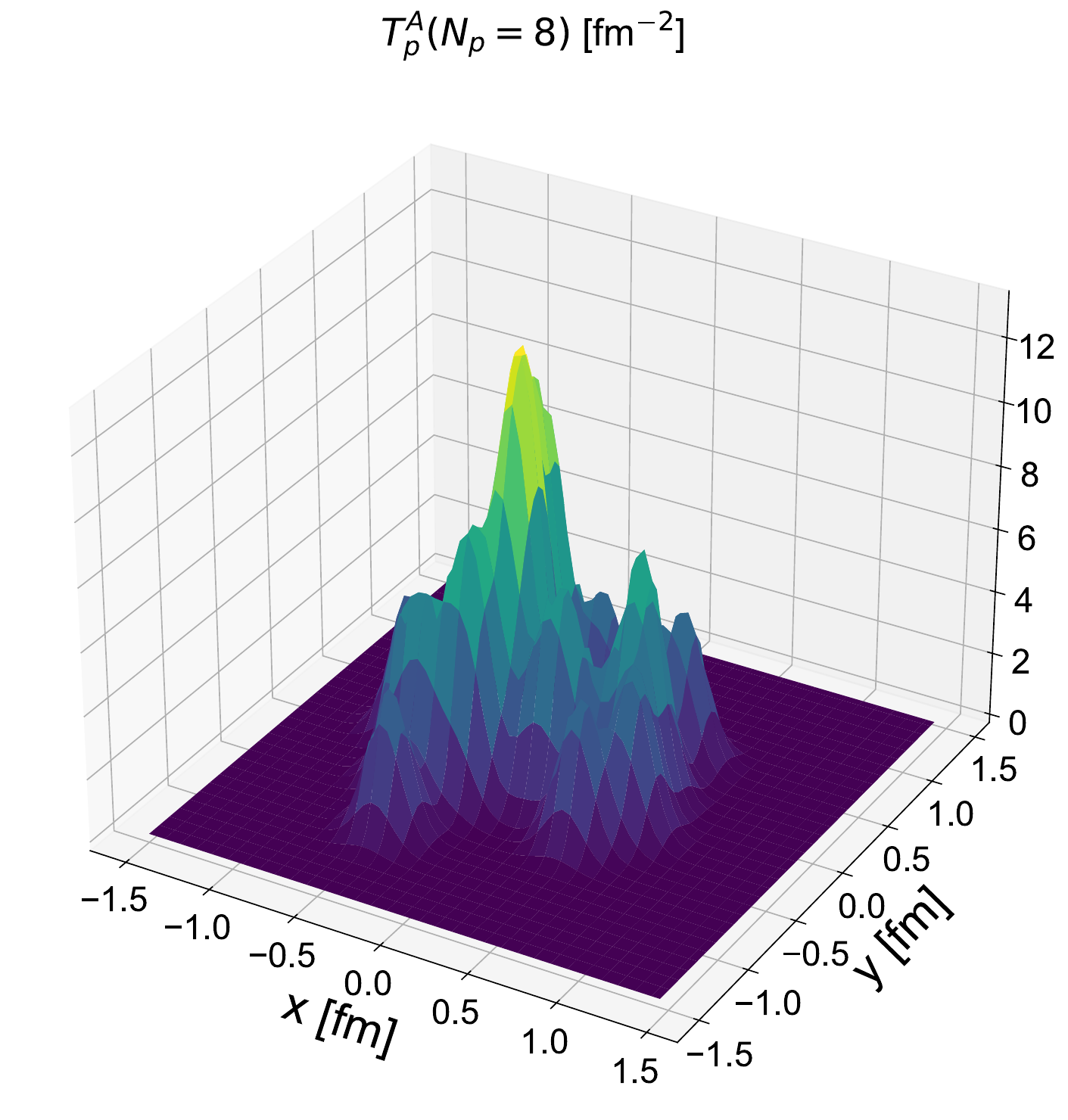}~~~~
\includegraphics[width=0.3\linewidth]{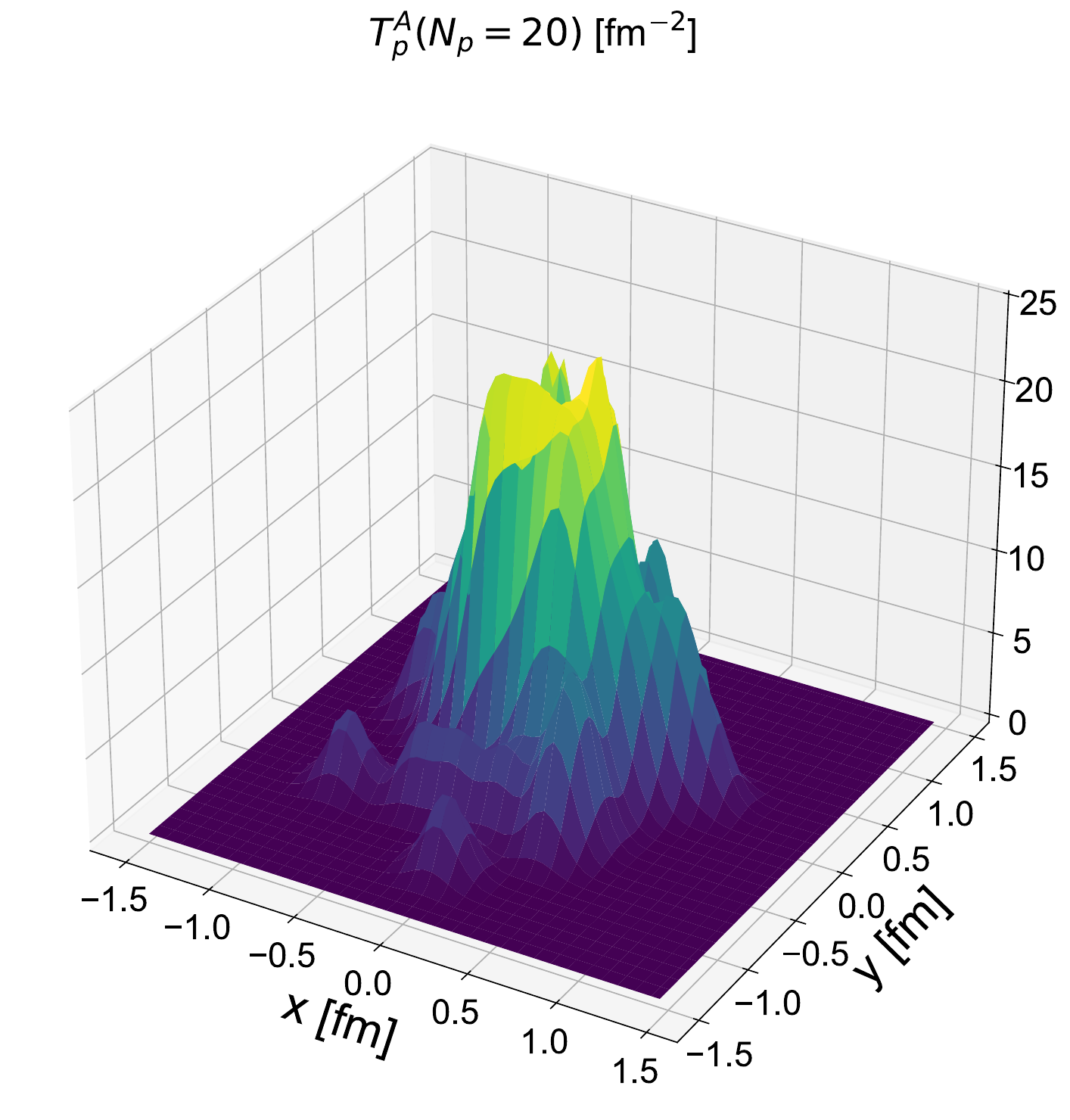}
    \caption{Thickness functions in fm$^{-2}$ used for the generation of the color
    charges. In the left panel we show $T_p$ for the proton
    (corresponding to $N_\mathrm{part}=1$),
    in the middle panel we show the case $N_\mathrm{part}=8$,
    corresponding to the average number of participants in 
    pA collisions at the LHC,
    and in the right panel we show 
$T_p$ for the case $N_\mathrm{part}=20$.}
    \label{Fig:tp1e2}
\end{figure*}

In Figure \ref{Fig:tp1e2} we show, for illustrative purposes, various charge profiles obtained with this model. Along with the profile of a proton, we also show the profiles we get for $N_\text{part}=8$ and $N_\text{part}=20$. 

\subsection{Wilson lines and gauge links}
\label{sec:gaugefields}
Once the charge has been generated, we know that the hard and the soft sectors are coupled via the Yang-Mills equations \cite{Yang:1954ek}
\begin{equation}
    D_\mu F^{\mu\nu}=J^\nu,
    \label{8.1}
\end{equation}
where $D_\mu=\partial_\mu-ig[A_\mu,\hspace{3pt} \cdot\hspace{3pt}]$ is the covariant derivative, $F^{\mu\nu}=\partial^\mu A^\nu-\partial^\nu A^\mu-ig[A^\mu,A^\nu]$ is the field strength tensor and $J^\mu$ the color current. The choice of the gauge and the symmetries of the system allow us to consider $A^-=A^i=0$, where $i=x,y$. The only component left is therefore $A^+\equiv \alpha$, which can be seen to obey the following Poisson equation
\begin{equation}
    \Delta_\perp \alpha(\mathbf{x}_\perp)=-\rho(\mathbf{x}_\perp).
    \label{9}
\end{equation}
In order to solve this equation, we Fourier-transform both sides of \eqref{9} in order to get
\begin{equation}
    \Tilde{k}^2_\perp \tilde{\alpha}_{n,k}^a=\tilde{\rho}_{n,k}^a.
    \label{9.1}
\end{equation}
The Fourier transformations are easily performed using the 
Fast Fourier Transform (FFT) package in Julia.
We get
\begin{equation}
\tilde{\alpha}_{n,k}^a=\frac{\tilde{\rho}_{n,k}^a}{\Tilde{k}_\perp^2+m^2}
    \label{9.4}
\end{equation}
where the discretized momentum $\Tilde{k}_\perp$ is given by
\begin{equation}
\Tilde{k}_\perp^2=\sum_{i=x,y}\left(\frac{2}{a_\perp}\right)^2\sin^2\left(\frac{k_ia_\perp}{2}\right),
    \label{9.2}
\end{equation}
and we introduced an infrared regulator $m=0.2$ GeV $=1$ fm$^{-1}$  acting as a screening mass. By then, Fourier-transforming back we get  $\alpha(\mathbf{x}_\perp)$ in coordinate space.

Once we obtained a solution for the gauge field in the covariant gauge, the Wilson line for each nucleus is evaluated as
\begin{equation}
    V_\mathbf{x}=\prod_{n=1}^{N_s}\exp\{ig\alpha_{n,\mathbf{x}}^a t^a\}
    \label{9.5}
\end{equation}
and then obtain the gauge links, $U$, for each of the two nuclei $A,B$ as
\begin{equation}
U_{\mathbf{x},i}^{A,B}=V_\mathbf{x}^{A,B}V_{\mathbf{x}+\hat{i}}^{\dagger,A,B}.
\label{9.6}
\end{equation}
In the above expressions $t^a$ are the $SU(3)$ group generators, i.e. the eight Gell-Mann matrices divided by 2. If the point $\mathbf{x}_\perp$ belongs to the edge of the simulation lattice, periodic boundary conditions have been implemented.

\subsection{Initialization of the boost-invariant fields}
At this point, we would like to derive the gauge link for the combined system of the two nuclei immediately after the collision. The gauge field will be the sum of each nucleus' contribution, but since QCD is a non-Abelian gauge theory, the resulting gauge link will not be the product of the gauge links of each nucleus. The total gauge link, $U_{\bm x,i}$, at the position
$\bm x$ in the transverse plane
is determined by solving a  set of eight equations, which are
\begin{equation}
    \text{Tr}[T_a(U_{\mathbf{x},i}^A+U_{\mathbf{x},i}^B)(\mathbb{I}+U_{\mathbf{x},i})-\text{h.c.}]=0,
    \label{10}
\end{equation}
with $a=1,\dots,N_c^2 - 1$.
In the case of  $N_c=2$ there is an exact solution of \eqref{10} for $U_{\mathbf{x},i}$ \cite{Krasnitz:1998ns}. 
Instead, for $SU(3)$   there is not exact solution and we have to solve \eqref{10} via an iterative method. Using this procedure we calculate the gauge links along the $x$ and $y$ direction, whereas for the longitudinal components we initialize $U_\eta=\mathbb{I}$.

We can also evaluate the color-electric fields, i.e. the canonical momenta with respect to the gauge links, at the initial time. 
The initial conditions on the lattice are implemented as~\cite{Fukushima:2011nq}
\begin{widetext}
\begin{align}
    &E_x=E_y=0,\nonumber\\
   &E^\eta=-\frac{i}{4ga_\perp^2}\sum_{i=x,y}\left[(U_i(\mathbf{x}_\perp)-\mathbb{I})(U_i^{B,\dagger}(\mathbf{x}_\perp)-U_i^{A,\dagger}(\mathbf{x}_\perp))+(U_i^\dagger(\mathbf{x}_\perp-\hat{i})-\mathbb{I})(U_i^{B}(\mathbf{x}_\perp-\hat{i})-U_i^{A}(\mathbf{x}_\perp-\hat{i}))-\mathrm{h.c.}\right].
   \label{11.2}
\end{align}
\end{widetext}

\subsection{Time evolution}
We now discuss the time evolution of the fields whose initialization
has been explained in the previous subsection. The electric fields are expressed in terms of link variables as \cite{Fukushima:2011nq}
\begin{align}
    \partial_\tau U_i(\mathbf{x})&=\frac{-iga_\perp}{\tau}E^i(\mathbf{x})U_i(\mathbf{x}),\label{eq:parisihatoltoilginger}\\
    \partial_\tau U_\eta(\mathbf{x})&=-iga_\eta\tau E^\eta(\mathbf{x}) U_\eta(\mathbf{x}).
    \label{11.3}
\end{align}
The term $a_\eta$ denotes the discretization step in the $\eta$-direction.

In order to reduce discretization errors in time, we implement a leapfrog algorithm, in which the electric fields and the gauge links evolve in staggered time steps. Specifically, the electric fields are defined at half-integer time steps, while the gauge links are defined at integer steps. At each iteration, the electric fields are updated first using the gauge links at integer time, and subsequently the gauge links are updated using the electric fields at half-integer time.

The equations of motion for the electric fields are discretized as:
\begin{widetext}
\begin{align}
E^i(\tau') &= E^i(\tau - \Delta \tau/2) + \Delta \tau \cdot \frac{i}{2ga_\eta^2 a_\perp\tau} \left[U_{\eta i}(\tau) + U_{-\eta i}(\tau) - \text{h.c.}\right] 
+ \Delta \tau \cdot \frac{i\tau}{2g a_\perp^3} \sum_{j\neq i} \left[U_{ji}(\tau) + U_{-ji}(\tau) - \text{h.c.}\right], \nonumber\\
E^\eta(\tau') &= E^\eta(\tau - \Delta \tau/2) + \Delta \tau \cdot \frac{i}{2g a_\eta a_\perp^2 \tau} \sum_{j=x,y} \left[U_{j\eta}(\tau) + U_{-j\eta}(\tau) - \text{h.c.}\right],
\label{12}
\end{align}
\end{widetext}
where \( \tau' = \tau + \Delta\tau/2 \), and the plaquette variables are defined as:
\begin{equation}
U_{\mu\nu}(\mathbf{x}) \equiv U_\mu(\mathbf{x})U_\nu (\mathbf{x}+\hat{\mu})U_\mu^\dagger(\mathbf{x}+\hat{\nu})U_\nu^\dagger(\mathbf{x}).
\label{13}
\end{equation}
Once the electric fields have been updated, the gauge links evolve via:
\begin{align}
U_i(\tau'')&=\exp\left[-\Delta \tau\cdot iga_\perp \frac{E^i(\tau')}{\tau'}\right]U_i(\tau),\label{11.4bisAAA}\\
U_\eta(\tau'')&=\exp\left[-\Delta \tau\cdot ig a_\eta \tau' E^\eta(\tau')\right]U_\eta(\tau),
\label{11.4}
\end{align}
where \( \tau'' = \tau + \Delta\tau \). The exponentiation of the electric fields ensures that the updated gauge links remain unitary matrices, as required for a consistent evolution in the $SU(3)$ gauge group.
In this scheme, the gauge links are initialized at time \( \tau_0 \) and evolve at integer multiples of the time step: \( \tau = \tau_0, \tau_0 + \Delta \tau, \tau_0 + 2\Delta \tau, \dots \). The electric fields, on the other hand, are defined at half-integer times: \( \tau = \tau_0 - \Delta \tau/2, \tau_0 + \Delta \tau/2, \tau_0 + 3\Delta \tau/2, \dots \)
It is important to note that the initial condition for the electric fields is assumed to be given at time \( \tau_0 - \Delta \tau/2 \), even if for practical purposes the available configuration is defined at \( \tau_0 \). This shift of half a time step is consistent with the leapfrog scheme and introduces only an \( \mathcal{O}(\Delta \tau^2) \) discretization error, which is within the overall accuracy of the algorithm. This convention allows one to begin the evolution without the need to explicitly compute a backward half-step.

The electric fields and gauge fields which we have dealt with so far fit into the definition of the longitudinal/transverse electric and magnetic components of the energy density. In particular, at time $\tau$ we have the electric components given by
\begin{align}
E_L^2(\tau)&\equiv E^{\eta}(\tau)\cdot E^{\eta}(\tau),\nonumber\\
E_T^2(\tau)&\equiv \frac{1}{\tau^2}[E^{x}(\tau)\cdot E^{x}(\tau)+E^{y}(\tau)\cdot E^{y}(\tau)].\label{13.1}
\end{align}
The magnetic components can be computed by tracing the plaquettes. In particular, 
\begin{align}
    B_L^2&=\frac{2}{g^2a_\perp^4}\text{Tr}(\mathbb{I}-U_{xy}),
\label{13.2_L}\\
    B_T^2&=\frac{2}{(ga_\eta a_\perp\tau)^2}\sum_{i=x,y}\text{Tr}(\mathbb{I}-U_{\eta i}).\label{13.2}
\end{align}
The above quantities fit into the definition of the energy density, which is $\varepsilon$ \cite{Fukushima:2011nq}
\begin{equation}
    \varepsilon=\langle \text{Tr}[E_L^2+B_L^2+E_T^2+B_T^2]\rangle.
    \label{13.3}
\end{equation}
The angular brackets $\langle \rangle$ denote ensemble average, numerically obtained by averaging the observables over the events.

\begin{figure}[t!]
    \centering
    \hspace{-0.8em}\includegraphics[width=0.98\linewidth]{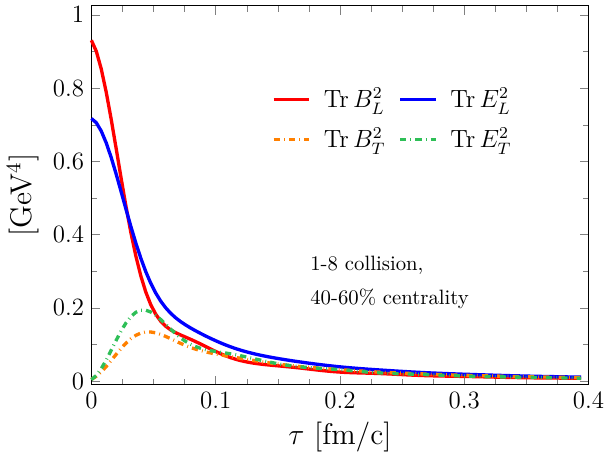}\\
    \vspace{1em}
    \includegraphics[width=0.95\linewidth]{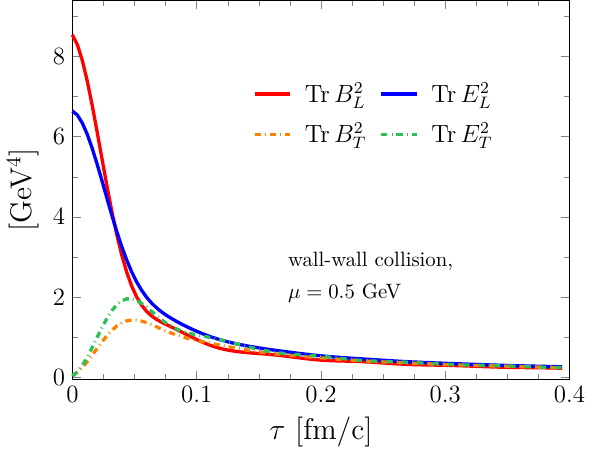}
    \caption{Color-electric and color-magnetic fields vs $\tau$ for pA collisions (1-to-8 participants collision, upper panel) and AA collisions (wall-wall collision, lower panel).}
    \label{Fig:campi}
\end{figure}

In Fig.~\ref{Fig:campi} we show the color electric and magnetic fields for pA (upper panel)
and AA (lower panel) collisions vs proper time $\tau$. The calculation for pA collisions has been performed as p-$N_{\text{part}}$ for $N_\text{part}=8$ (corresponding to 40-60$\%$ centrality class, check \cite{ALICE:2014xsp}), while for the AA ‘wall-wall' collision we have fixed $\mu=0.5$ GeV. In obtaining this plot, and for all the calculations in this paper, the transverse size has been fixed as $L_\perp=3$ fm, the number of lattice points is $N_\perp=64$ and we fix $N_s=50$. From these plots one can infer general features of the glasma: at $\tau=0^+$ only non-zero components are the longitudinal ones, then as time passes also the transverse components develop. At late times, all the fields are weak, and follow a common trend as $\tau^{-1}$.

\section{Heavy quark dynamics in the pre-equilibrium stage}
\label{sec:heavy_quark_dynamics}

In this section, we discuss how we model the diffusion of the 
heavy quarks in the pre-equilibrium stage.
Following~\cite{Pooja:2024rnn, Oliva:2024rex}, we model 
the dynamics of the $c$ and $b$ quarks using semi-classical equations of motion,
namely the Wong equations. 
They describe the time evolution of the coordinate {\bf x}, 
momentum {\bf p}, and color charge $Q_a$ (with
$a=1,\dots,N_c^2-1$) of the heavy quarks~\cite{Wong:1970fu, Heinz:1984yq}.
In the laboratory frame, the Wong equations read
\begin{eqnarray}
    \dfrac{\mathrm{d}x^i}{\mathrm{d}t} & =&\frac{p^i}{E},
    \label{eq:wong1}\\
    \dfrac{\mathrm{d}p^i}{\mathrm{d}t} & =&gQ^aF^{i\mu,a}\frac{p_\mu}{E},
    \label{eq:wong2}\\
    \dfrac{\mathrm{d}Q_a}{\mathrm{d}t} & =&-gf^{abc} A_\mu^bQ_c\frac{p^\mu}{E},
    \label{eq:wong3}
\end{eqnarray}
where $i=x,y,z$ and $\mu=0,\dots,3$. 
Moreover, $f^{abc}$ are the structure constants of the gauge group $SU(N_c)$ and $E=\sqrt{ \bm p^2+M^2}$ is the energy of the quark with mass $M$. 
The Wong equations have been recently used to study the dynamics of the heavy quarks in the pre-equilibrium stage of high-energy nuclear collisions; see \cite{Ruggieri:2018ies,Ruggieri:2018rzi,Sun:2019fud,Liu:2019lac,Jamal:2020fxo,Liu:2020cpj,Sun:2021req,Khowal:2021zoo, Pooja:2022ojj,Pooja:2024rnn, Avramescu:2024poa, Avramescu:2024xts} and references therein.

We solve Eqs.~\eqref{eq:wong1}-\eqref{eq:wong3} in the background of the evolving glasma fields. The gauge field $A_\mu$ 
in Eq.~\eqref{eq:wong3}
is determined by (the $A_\tau$ component is zero because of the gauge choice)
\begin{align}
    A_{i}(\mathbf{x})&=-\frac{1}{2ig a_\perp}\left\{\delta U_i(\mathbf{x})-\frac{1}{3}\left[\mathrm{Tr}\,\delta U_i(\mathbf{x})\right]\mathbb{I}\right\},
    \label{eq:gauge_field_xy}\\
    A_\eta(\mathbf{x})&=-\frac{1}{2ig a_\eta}\left\{\delta U_\eta(\mathbf{x})-\frac{1}{3}\left[\mathrm{Tr}\,\delta U_\eta(\mathbf{x})\right]\mathbb{I}\right\}.\label{eq:gauge_field_eta}
\end{align}
where 
$i=x,y$, and we define
$\delta U\equiv U-U^\dagger$. The color-electric and color-magnetic fields are defined in terms of the field strength tensor through
\begin{eqnarray}
&&F_{\tau i}=E_i,~~~F_{\eta i}=\tau\epsilon_{ij}B_j,\\
&&F_{\tau\eta} = \tau E_\eta,~~~F_{xy}=-B_\eta,\label{eq:EBfields}
\end{eqnarray}
where $F_{\mu\nu}$ is extracted from the
plaquettes as 
\begin{equation}
F_{\mu\nu} = \frac{1}{2i g a_\mu a_\nu}
\left\{\delta U_{\mu\nu}
-
\frac{1}{3}\left[\mathrm{Tr}\,\delta U_{\mu\nu}\right]\mathbb{I}
\right\}.
\label{eq:definizioneFmunu}
\end{equation}
Equations \eqref{eq:wong1} and \eqref{eq:wong2} have been solved using the Euler method. Instead, Eq. \eqref{eq:wong3} has been solved by exponentiation, i.e. evolution from one timestep to another is performed for each HQ as
\begin{equation}
    Q(\tau+\Delta \tau)=\mathcal{U}^\dagger(\tau)Q(\tau)\,\mathcal{U}(\tau)
    \label{eq:exponentiation_charge_evolution}
\end{equation}
where
\begin{equation}
    \mathcal{U}(\tau)=\exp\left[i\,(A_xv_x+A_yv_y)\,\Delta \tau\right]
    \label{eq:gauge_exponential}
\end{equation}
and $v_i=p_i/E$ is the $i^\text{th}$ component of the velocity of the HQ here considered. This approach preserves the Casimir invariants which follow from the evolution of the color charges governed by Eq.~\eqref{eq:wong3}. 
For $N_c=3$ the conserved Casimir invariants are
\begin{equation}
q_2 = Q_a Q_a,~~~q_3=d_{abc}Q_a Q_b Q_c,
\label{eq:mostrasumarte}
\end{equation} 
with
\begin{equation}
q_2=\frac{N_c^2-1}{2} ~~\text{and}~~q_3=\frac{(N_c^2-4)(N_c^2-1)}{4N_c}, \label{eq:casimir_values}
\end{equation}
and $d_{abc}$ are the symmetric structure constants \cite{Avramescu:2023qvv}.

Within our model, 
we follow previous works and neglect the back-reaction of the quarks 
in the Yang-Mills equations: it has been shown~\cite{Liu:2020cpj} that 
such back-reaction 
does not substantially affect the diffusion of heavy quarks
within the lifetime of the pre-equilibrium stage.\\

As far as the initialization is concerned, in this work we consider the heavy quarks produced at mid-rapidity, so that the longitudinal coordinate of the heavy quarks is kept $z=0$. In the transverse plane, the initial positions of the heavy quarks are extracted on an event-by-event basis: for each event, we extract the position of each heavy quark from a probability distribution given by the energy density in the transverse plane \cite{Oliva:2024rex}. 
The transverse momenta of the heavy quarks are initialized according to the Fixed Order + Next-to-Leading Log (FONLL) QCD result that reproduces the $D$-mesons and $B$-mesons spectra in $pp$ collisions after fragmentation~\cite{Cacciari:1998it, Cacciari:2001td, Cacciari:2012ny}, parametrized as \cite{Pooja:2024rnn}:
\begin{equation}
    \left.\frac{dN}{d^2 p_T}\right|_\mathrm{prompt} = \frac{x_0}{\big(1 + x_3\cdot {p_T}^{x_1}\big)^{x_2}},\label{eq:FONLL}
\end{equation}
with parameters $x_0=20.2837$, $x_1=1.95061$, $x_2=3.13695$, $x_3=0.0751663$ for charm quarks. 
The azimuthal angle $\phi$ of each heavy-quark momentum is chosen randomly between 0 and $2\pi$, so the initial distribution is isotropic in momentum space. Moreover, we assume initially $p_z=0$ for quarks and their companion antiquarks. Finally, let us deal with the initialization of the color charges, which needs to satisfy the constraints~\eqref{eq:mostrasumarte}.
In order to do so, we firstly 
fix the charge of one heavy quark 
in the ensemble
to be~\cite{Avramescu:2023qvv}
\begin{equation}
Q_{a}^0=-1.695\,\delta_{a5}-1.062\,\delta_{a8},
    \label{eq:HQcharge0}
\end{equation}
which explicitly satisfies 
the conditions~\eqref{eq:mostrasumarte}, and then generate the charges of all the other heavy quarks as
\begin{equation}
Q_a=Q_{b}^0 U^{ab}, ~~~U^{ab}=2\,\text{Tr}[T^a U T^b U^\dagger],
    \label{eq:HQ_charge_all}
\end{equation}
where $U$ is an $SU(3)$ random matrix which is extracted for each heavy quark according to the Haar measure. To finish this section, the formation time of the heavy quarks has been fixed as $\tau_{\text{form}}=1/2M$, being $M$ the heavy quark mass: for the charm quark, this corresponds to $\tau_{\text{form}}=0.077$ fm/c.

\section{\texorpdfstring{$R_{pA}$}{RpA} of heavy quarks}
\begin{figure}[t!]
\centering
\includegraphics[width=\linewidth]{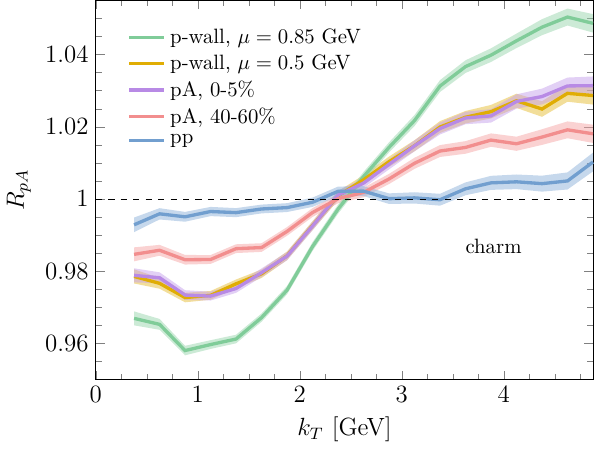}
\caption{Nuclear modification factor $R_{pA}$ defined in \eqref{eq:RpA_definition}, computed at $\tau=0.4$ fm/c,
for several collision systems.}
\label{Fig:RpA_charm_pWall_pA_pp}
\end{figure}

The first effect we here want to highlight is the modification of the HQ spectrum due to the action of the glasma fields. In particular, let us define the nuclear modification factor $R_{pA}$ as the ratio of the HQ spectrum at a time $\tau$ over the HQ spectrum at initial time:
\begin{equation}
    R_{pA}(k_T)=\frac{\left.dN/d^2\bm{k}_T(k_T)\right|_{\tau}}{\left.dN/d^2\bm{k}_T(k_T)\right|_{0}}.
    \label{eq:RpA_definition}
\end{equation}
We show our results for $R_{pA}$ vs $k_T$ in Fig. \ref{Fig:RpA_charm_pWall_pA_pp}, for various collision systems, at $\tau=0.4$ fm/c: we take this time as a representative value of the
proper time at which the switch from our picture to one based on 
hydro or kinetic theory takes place. In particular, we show the $R_{pA}$ obtained for ‘p-wall' collisions (for two different values of the MV parameter, green $\mu=0.85$ GeV and yellow $\mu=0.5$ GeV), as well as for pA collisions at 0-5$\%$ and 40-60$\%$ centralities (purple and red curves respectively), and pp collisions (blue curve). The aforementioned centralities are reproduced as p-$N_\mathrm{part}$ collisions with $N_\mathrm{part}=16$ and $N_\mathrm{part}=8$, respectively (check \cite{ALICE:2014xsp}). Along with the data, which represent the ensemble average, we also show error bands corresponding to the statistical uncertainty over the mean. Here and for all the numerical data which follow, we show results obtained with 1500 events.

Notice that in all cases we observe a smaller-than-one value for $R_{pA}$ for $k_T\lesssim 2.5$ GeV, while for $k_T\gtrsim 2.5$ GeV we observe that $R_{pA}$ is greater than 1. This goes to show that the action of the glasma on the spectrum is a net injection of energy, leading to a depletion of the spectrum at low momenta in favor of the higher momenta. A striking feature that we observe is that all the curves share a common value of transverse momentum at which the spectrum of the charm quarks is not modified by the action of the glasma fields, and we are able to locate such point around $k_T= 2.5$ GeV. 

Moving on to the differences among curves, we observe that the maximum deviation from the value $R_{pA}=1$ is achieved in the p-wall data at $\mu=0.85$ GeV (green curve), followed by $\mu=0.5$ GeV (yellow curve), which show a maximum displacement of $R_{pA}$ from 1 of around $5\%$ and $3\%$ in the $k_T$ range of reference, respectively. Unsurprisingly, the charm quarks experience a greater amount of spectrum modification in glasma systems with greater energy, or equivalently, with greater saturation scale $Q_s$. If we consider the subnucleonic structure, we observe that the lower the number of participants, the smaller the modification of the momentum spectrum.  This can be interpreted once again, in terms of energy density, since a smaller number of participant nucleons implies a lower energy of the glasma, therefore a reduced modification of the initial HQ spectrum. The comparison among the ‘p-wall' results and the ‘p-nucleons' results shows that, interestingly, the p-wall data at $\mu=0.5$ GeV (yellow) are compatible with the pA data at 0-5$\%$ centrality (purple), indicating that the energy densities of the color fields in the two systems are comparable. Finally, we note that our results are in qualitative agreement with the ones from \cite{Ruggieri:2018rzi}; quantitative comparisons cannot however be made, since the latter work simulates the glasma fields on a static box, while here we allow the full expansion of the gluon fields. Other works have investigated the spectrum modification of charm quarks in AA collisions \cite{Liu:2020cpj, Avramescu:2024poa, Avramescu:2024xts}, finding a more pronounced effect than ours, as expected since in nucleus-nucleus collisions the glasma fields are more intense.

It is worth considering that, for all the systems considered, the spectrum modification of charm quarks is limited to a few percent, reaching at most a $5\%$ deviation with respect to the value 1. Such a value is much smaller than the ones observed experimentally for charmed hadrons produced in proton-lead collisions, see e.g. \cite{LHCb:2022dmh}. This goes to show that, in pA collisions at LHC, most of the observed suppression of the transverse momentum spectrum is due to the later stages of the nuclear collision, while the initial stages have only a mild effect. In the following section we will see that, on the other hand, the effect that the initial stages have on the observed momentum anisotropies of the charm quarks is quite substantial.

\section{Elliptic flow \texorpdfstring{$v_2$}{v2}}
\label{sec:anisotropic_flows}
Evolution of HQs in the early stage does affect not only their spectrum, but their entire distribution in
momentum space. In particular,
momentum anisotropies of the glasma color fields can be transferred to
HQs.
In this section,
we analyze in some detail the elliptic flow, $v_2$,
of HQs at the end of their evolution in the early stage. 
Particularly,
we focus on 
$v_2$ computed
both with the event plane  
and the 2-particle correlation methods.
Our main goal is to give a quantitative estimate
of the $v_2$ of HQs at the end of their evolution in the
early stage, both for pA and pp collisions.


\subsection{Event-Plane (EP) method}
\label{subsec:EP_method}

A first method for evaluating the $v_n$ is what is called the \emph{event-plane} (EP) method. It consists in evaluating the event plane angle $\Psi_n$ as a weighted average over the particle spectrum. In particular, we identify $\Psi_n$ as:
\begin{equation}
\Psi_n = \frac{1}{n}\arctan\frac{S_n}{C_n},
\label{eq:angoloPsin}
\end{equation}
with
\begin{align}
C_n &= \int \cos (n\phi) \frac{dN}{d^2\bm{k}_T} \, d^2\bm{k}_T, \\
S_n &= \int \sin (n\phi) \frac{dN}{d^2\bm{k}_T} \, d^2\bm{k}_T,
\end{align}
being $\phi$ the azimuthal angle of $\bm{k}_T$. In other words, the event plane angle is oriented towards the direction in momentum space with largest particle yield. From this, the anisotropic flow $v_n\{\mathrm{EP}\}$ at for a given momentum modulus $k_T$ is defined by integrating over all $\phi$ at fixed $k_T$:
\begin{equation}
v_n\{\mathrm{EP}\}(k_T) =\frac{\displaystyle{\int d\phi\, \frac{dN}{d^2\bm{k}_T}\, \cos[n(\phi - \Psi_n)]}}
{\displaystyle{\int d\phi\, \frac{dN}{d^2\bm{k}_T}}}.
\label{eq:ep_vn}
\end{equation}
The final result is obtained via an average over the events.

\subsection{Two-Particle Correlation (2PC) method}
\label{subsec:2PC_method}

An alternative to the event-plane method for the extraction of anisotropic flow coefficients $v_{n}$ is provided by the so-called \emph{two-particle correlation} (2PC) method, which is based on experimental grounds. Differently from before, the basic idea here does not consist in explicitly calculating the event plane angle $\Psi_n$, but one rather indirectly infers its existence by studying the azimuthal correlation between pairs of particles. More specifically, for each $k_T$ we compare particles belonging to the bin centered in $k_T$ with particles in a broader ‘‘reference'' momentum region. From this information, we are able to derive the anisotropic flows.\\

To determine the desired harmonic $v_n$, for a given event we start by defining the quantity $Q_n$ as
\begin{equation}
Q_{n} = \sum_{j=1}^{M} w_{j}\, e^{i n \phi_{j}},
\end{equation}
where the sum runs over the $M$ particles, each with azimuthal angle $\phi_{j}$ and weight $w_{j}$. In our analysis the weights can be taken either from the particle spectrum $dN/d^2\bm{k}_T$ or, equivalently, set to $w_{j}=1$ if one wants to perform a pure particle count. Assuming for simplicity that $w_{j}=1$, let $N_{j}$ be the number of particles having momenta included in the bin centered at $k_{T}^{j}$, where $j$ runs over the number of bins. From these particles we can construct
\begin{equation}
    Q_{n}^{(j)} = \sum_{a=1}^{N_j} e^{i n \phi_{j,a}}.
\end{equation}
Now let us consider the so-called \emph{reference set}, defined by all particles with $k_{T}$ within a fixed interval $[k_{T}^{\mathrm{min}}, k_{T}^{\mathrm{max}}]$. Let the number of particles in such set be $N_{\mathrm{ref}}$, the corresponding $Q_{n}^{\mathrm{ref}}$ is defined as
\begin{equation}
    Q_{n}^{\mathrm{ref}} = \sum_{j\in \mathrm{ref}} Q_n^{(j)}=\sum_{j \in \mathrm{ref}} \sum_{a=1}^{N_j} e^{i n \phi_{j,a}}.
\end{equation}
The first sum is intended over the bins included in the reference interval. In this work the reference interval has been chosen as $[1, 6]~\text{GeV}$.

The basic two-particle correlator between the bin of interest and the reference set is defined as
\begin{equation}
V_{n}^{\Delta}(k_{T}^{j}, \mathrm{ref}) = 
\frac{\text{Re}\left[Q_{n}^{(j)} \, Q_{n}^{\mathrm{ref}*} \right]}{N_{j}\,N_{\mathrm{ref}}} 
\label{eq:v22pc},
\end{equation}
where $N_{j}\,N_{\mathrm{ref}}$ indicates the number of all possible particle pairs. Similarly, one can define the reference-reference correlator as
\begin{equation}
V_{n}^{\Delta}(\mathrm{ref}, \mathrm{ref}) = 
\frac{\text{Re}\left[ Q_{n}^{\mathrm{ref}} \, Q_{n}^{\mathrm{ref}*} \right]}{N_{\mathrm{ref}}^{2}} 
= \frac{|Q_{n}^{\mathrm{ref}}|^{2}}{N_{\mathrm{ref}}^{2}}.
\end{equation}

Our purpose is to consider correlations among particle pairs. However, the definition above may include trivial ``self-correlation'' contributions when the bin of interest $k_{T}^{j}$ lies inside the reference interval. In such case, the same particle can appear both in $Q_{n}^{(j)}$ and in $Q_{n}^{\mathrm{ref}}$, producing additional spurious terms. To correct for this, one explicitly subtracts the self-correlation contributions, so to take into account, for each particle in the bin $j$, the correlation with all reference particles except itself. This leads to the modified expression
\begin{equation}
V_{n}^{\Delta}(k_{T}^{j}, \mathrm{ref}) = \frac{ \text{Re} \left[ Q_{n}^{(j)} Q_{n}^{\mathrm{ref}*} \right] - N_{j} }{ N_{j} (N_{\mathrm{ref}}-1)},
\end{equation}
valid if the bin $j$ lies inside the reference interval. On the other hand, if the bin of interest lies \emph{outside} the reference range, no particle is double-counted, and one should simply use the previously given definition \cref{eq:v22pc}. Also the reference-reference correlator can be corrected analogously by excluding the trivial self-pairs, yielding
\begin{equation}
V_{n}^{\Delta}(\mathrm{ref}, \mathrm{ref}) =
\frac{|Q_{n}^{\mathrm{ref}}|^{2} - N_{\mathrm{ref}}}{ N_{\mathrm{ref}}(N_{\mathrm{ref}}-1)}.
\end{equation}

Given all of this, the 2PC estimate of the anisotropic flow coefficient for a given transverse momentum bin $k_T^j$ is given by
\begin{equation}
v_{n}\{\mathrm{2PC}\}(k_{T}^{j}) = 
\frac{ V_{n}^{\Delta}(k_{T}^{j}, \mathrm{ref}) }{ \sqrt{ V_{n}^{\Delta}(\mathrm{ref}, \mathrm{ref})}}.
\end{equation}
The final result is obtained via an average over the events.

\subsection{Results: glasma \texorpdfstring{$v_2$}{v2}}
To begin with,
we briefly discuss our results for the 
elliptic flow of the evolved glasma fields. 
Results for $v_2$ of the early-stage fields 
for both AA and pA collisions
are known in the literature~\cite{Schenke:2015aqa},
hence we show them just to ensure that our calculation
produces results in agreement with those of~\cite{Schenke:2015aqa}.
In order to define a $v_2$ for the glasma, we firstly have to
extract the gluon distribution function from the fields.
In our work
we follow~\cite{Lappi:2009xa} and define the gluon spectrum as
\begin{widetext}
\begin{align}
&\frac{\d N}{\d^2 \bm{k}_\perp \d y} =\nonumber\\
&\frac{1}{(2 \pi)^2}\Tr \bigg\{
\frac{1}{\tau |\bm{k}_\perp|} E^i(\bm{k}_\perp)E^i(-\bm{k}_\perp) 
+ \tau |\bm{k}_\perp|  A_i(\bm{k}_\perp)A_i(-\bm{k}_\perp) + \frac{1}{|\bm{k}_\perp|} \tau E^\eta(\bm{k}_\perp) E^\eta(-\bm{k}_\perp)
+ \frac{|\bm{k}_\perp|}{\tau} A_\eta(\bm{k}_\perp) A_\eta(-\bm{k}_\perp)
\nonumber \\
&~~~~~~~~~~~~~~+ i \Big[ E^i(\bm{k}_\perp) A_i(-\bm{k}_\perp) - A_i(\bm{k}_\perp)E^i(-\bm{k}_\perp) \Big]+i \Big[ E^\eta(\bm{k}_\perp) A_\eta(-\bm{k}_\perp) -  A_\eta(\bm{k}_\perp) E^\eta(-\bm{k}_\perp) 
\Big]
\bigg\},\label{eq:second_line_spectrum_text}
\end{align}
\end{widetext}
where the fields are evaluated in the 2-dimensional Coulomb gauge. Notice that, while the first line of \eqref{eq:second_line_spectrum_text} possesses reflection symmetry, the interference terms in the second line are odd under the transformation $\bm{k}_\perp \to -\bm{k}_\perp$. These do not contribute when the gluon spectrum is averaged over the angle of $\bm{k}_\perp$, integrated over $\bm{k}_\perp$ or averaged over configurations of the sources. Neglecting them, as for instance done in \cite{Lappi:2003bi}, would be justified if one was interested in evaluating the single inclusive gluon spectrum. In the case of two-gluon correlations they cannot, however, be ignored.
 Note that the expression in Eq. \eqref{eq:second_line_spectrum_text} acts, as just discussed, as a proxy for a particle distribution, and it is not the only possibility considered in the literature, cfr. \cite{Lappi:2003bi, Jia:2022awu, Lappi:2009xa, Schenke:2015aqa, Berges:2013eia}.\\

\begin{figure}[t!]
\centering
\includegraphics[width=\linewidth]{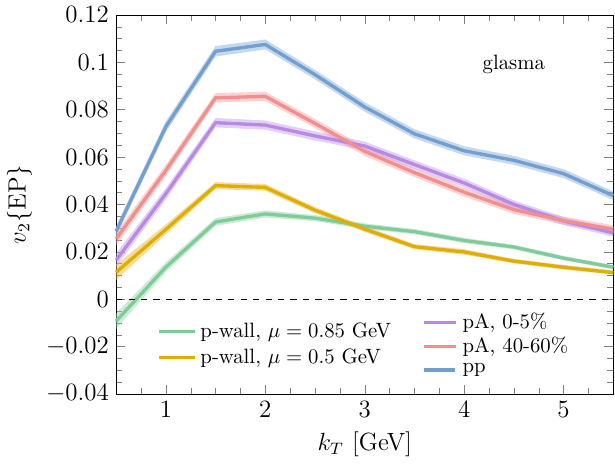}
\caption{Elliptic flow $v_2$ of the glasma fields versus $k_T$, computed at 
$\tau=0.4$ fm/c,
for several collision systems.}
\label{Fig:v2_glasma}
\end{figure}

In Fig. \ref{Fig:v2_glasma} we plot 
$v_{2}\{\mathrm{EP}\}$ versus $k_T$ for several collision systems,
computed at $\tau=0.4$ fm/c. 
In particular, we show the elliptic flow obtained for 'p-wall' collisions for two different values of the MV parameter $\mu=0.85,0.5$ GeV (green and yellow curves, respectively), as well as for pA collisions at 0-5$\%$ and 40-60$\%$ centralities (purple and red), and pp collisions (blue). 
As already mentioned, 0-5$\%$ and 40-60$\%$ centralities correspond 
to p-$N_\mathrm{part}$ collisions with $N_\mathrm{part}=16$ and $N_\mathrm{part}=8$, respectively (check \cite{ALICE:2014xsp}).

We notice that
all curves 
in Fig.~\ref{Fig:v2_glasma}
show similar qualitative behavior, reaching a maximum for a transverse momentum of around $k_T\sim 2$ GeV and then decreasing as $k_T$ increases. 
Within the p-walls calculations,
we find that
$v_2$ for $\mu=0.85$ GeV
is lower than the one for
$\mu=0.5$ GeV: this is related to the smaller size
of the correlation  
domains in the former case, which 
implies that 
a larger number of randomly oriented color domains 
are present in the interaction region and, as a consequence,
anisotropy is washed out when taking the integral over the whole
transverse plane. Moving on to the results corresponding to the p-$N_\mathrm{part}$ cases, we observe that for a smaller number of participants, the elliptic flow is higher. For the proton-proton case, in particular, the $v_2$ is quite significant, with a maximum of around $v_2\sim 0.11$. This ordering with the number of participants is straightforward to interpret, since a lower number of participants implies a lower average saturation scale $Q_s$. Following the previous argument, this in turn implies an increase in the typical size of the correlation domains, hence to a larger anisotropy of the system. In other words, when the number of participants is higher the distribution of the correlation domains is more isotropic in the transverse plane, resulting on average in a smaller
anisotropy. Finally, we note that our data for $v_2(k_T)$ in pA collisions are in agreement with the ones from \cite{Schenke:2015aqa}.

\begin{figure}[t]
\centering
\includegraphics[width=\linewidth]{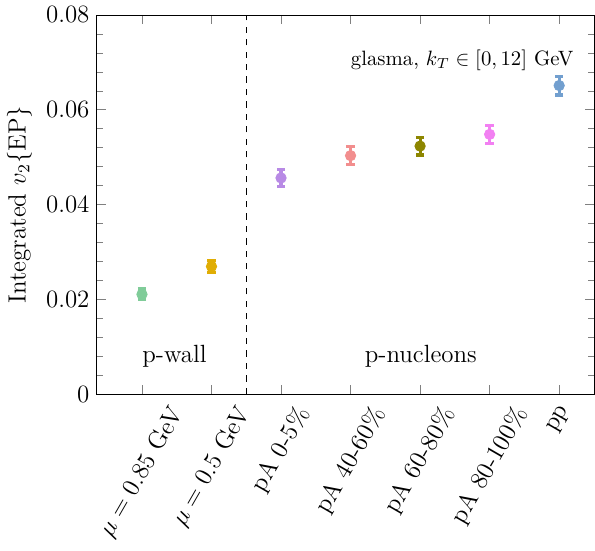}
\caption{Elliptic flow of glasma fields, computed at $\tau=0.4$ fm/c for several collision systems, integrated over the $k_T$ range $[0,12]$ GeV.}
\label{Fig:v2_glasma_integrated}
\end{figure}

In Fig.~\ref{Fig:v2_glasma_integrated} we plot the
integrated $v_2$ of the gluon fields at $\tau=0.4$ fm/c for several centrality classes of pA collisions, as well as for the two p-wall models, shown 
in Fig.~\ref{Fig:v2_glasma}. By ‘integrated' we refer to an averaging procedure of the $v_2$ over the momentum spectrum, within the range $[0,12]$ GeV. The same considerations that we have previously made still hold.

\subsection{Results: charm quark $v_2$}
The main topic of our study is the $v_2$ of charm quarks produced, in the early stages, via the interaction with the evolving glasma fields. In fact
the momentum anisotropy that the glasma possesses, due to the domain-like structure of the fields in the transverse plane, gets transmitted to HQs in the early stage.

\begin{figure}[t!]
\centering
\includegraphics[width=\linewidth]{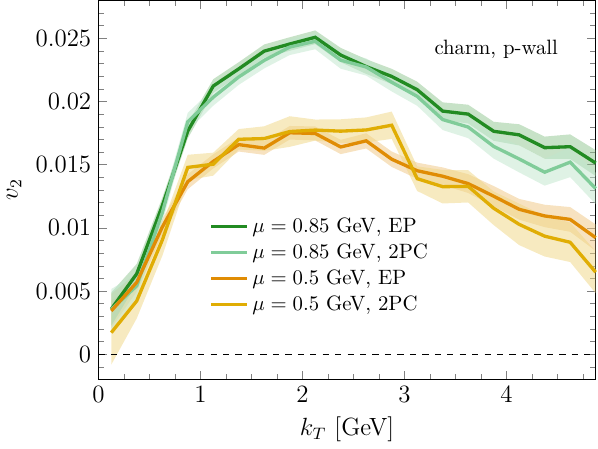}
\caption{Elliptic flow $v_2$ of charm quarks in p-wall collisions, versus $k_T$, computed at $\tau=0.4$ fm/c. We compare the results obtained for two different values of $\mu$, and with two different methods, Event-Plane (EP) and 2-Particle Correlations (2PC).}
\label{Fig:v2_charm_pWall_EPvs2PC}
\end{figure}

We begin our investigation by 
comparing the two methods for the calculation of $v_2$ that we have outlined in sections \ref{subsec:EP_method} and \ref{subsec:2PC_method}, namely EP and 2PC. 
As explained in Sec. \ref{sec:heavy_quark_dynamics}, we initialize the heavy quarks in momentum space with 
an isotropic distribution, hence the anisotropies measured at the
end of our evolution are solely due to the interaction with the
gluon fields. In Fig.~\ref{Fig:v2_charm_pWall_EPvs2PC}, we compare $v_2$ 
of charm quarks, computed within the 'p-wall' model of pA collisions,
at $\tau=0.4$ fm/c. 
In particular, we compare the results obtained using EP and 2PC for two values of the MV parameter. We perform 1500 events, each simulating the motion of 20000 charm quarks.\footnote{The heavy quarks used in our simulations serve as test particles, since we only evaluate ensemble-averaged quantities. Their number does not correspond to the number of actual HQs produced in the collisions.} We notice that the extracted $v_{2}$ values from 2PC are in very good agreement with the EP results, each lying within the error bars of the other. This holds for both the $\mu=0.5$ GeV and the $\mu=0.85$ GeV results, showing that the observed azimuthal anisotropies are physical in origin and not artifacts of a particular analysis technique, either EP or 2PC. Moving on to the physical features
shown in Fig.~\ref{Fig:v2_charm_pWall_EPvs2PC}, 
we find that for both values of $\mu$ the $v_2$ reaches a maximum value for $k_T\sim 2$ GeV, after which we have a decreasing behavior. This closely resembles the generic trends seen in heavy-ion phenomenology, i.e. that elliptic flow is typically most pronounced at intermediate transverse momenta before flattening at higher values. The maximum value of elliptic flow is found to be around $v_2\sim$ 0.025 for $\mu=0.85$ GeV (green curves) and $v_2\sim$ 0.018 for $\mu=0.5$ GeV (orange curves), so at higher $\mu$ we observe higher values of $v_2$. This can be interpreted by noting that at higher $\mu$ the glasma fields have a greater energy density, therefore the color electric and magnetic fields in the initial stages are more efficient in transmitting their momentum anisotropy to the heavy quarks which evolve therein.

\begin{figure}[t!]
\centering
\includegraphics[width=\linewidth]{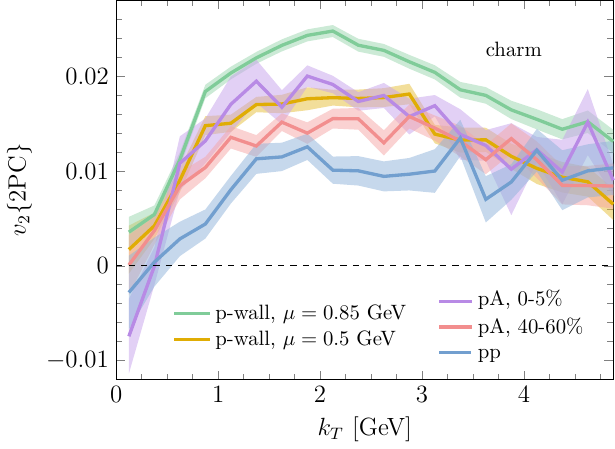}
\caption{Elliptic flow $v_2$ of charm quarks versus $k_T$, computed at $\tau=0.4$ fm/c, for several collision systems.}
\label{Fig:v2_charm_pWall_pA_pp}
\end{figure}

In Figure \ref{Fig:v2_charm_pWall_pA_pp} we show the results of the elliptic flow $v_2$ vs $k_T$ for various collision systems using the 2PC method: along with the p-wall data (the same as in Fig. \ref{Fig:v2_charm_pWall_EPvs2PC}), we also show results for pA collisions at 0-5$\%$ and 40-60$\%$ centralities (purple and red curves), as well as pp collisions (blue curve). A striking feature is that all the curves follow the same qualitative behavior, reaching a maximum at around $k_T\sim 2$ GeV and then decreasing for higher $k_T$. Quantitatively, we observe that the smaller the number of participants, the lower the obtained $v_2$ for charm quarks:  once again, a smaller number of participant nucleons implies a lower strength of the gluon fields, hence the anisotropy is transmitted less efficiently to HQs. It is interesting to notice that the ordering of the $v_2$ with respect to the number of participants is opposite to the one we have observed in Fig. \ref{Fig:v2_glasma}. Moreover, we also note that the results for pA collisions at 0-5$\%$ centrality (purple curve) are comparable, both qualitatively and quantitatively, with the p-wall results at $\mu=0.5$ GeV (yellow), even though the first system is manifestly more anisotropic than the second (because of the presence of a subnucleonic structure). We can conclude that, in this comparison, the greater spatial anisotropy which is observed in pA collisions at 0-5$\%$ is not transmitted to momentum as efficiently as it is in the p-wall case. This leads to a balancing effect which renders the yellow and the purple curves compatible with each other, within error bars.

\begin{figure}[t]
\centering
\includegraphics[width=\linewidth]{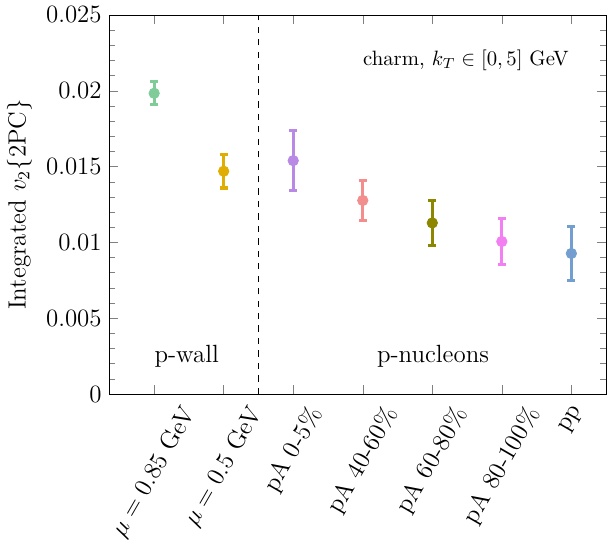}
\caption{Elliptic flow of charm quarks, computed at $\tau=0.4$ fm/c for several collision systems, integrated over the $k_T$ range $[0,5]$ GeV.}
\label{Fig:v2_charm_integrated}
\end{figure}

These conclusions are supported also in Fig. \ref{Fig:v2_charm_integrated}, in which we show the integrated elliptic flow of charm quarks at $\tau=0.4$ fm/c. Along with the same cases we have shown in the previous plot, here we also show results for the integrated $v_2$ obtained in pA collisions at 60-80$\%$ and 80-100$\%$ centrality classes, corresponding to p-$N_\mathrm{part}$ collisions with $N_\mathrm{part}=5$ and $N_\mathrm{part}=3$, respectively (check once again \cite{ALICE:2014xsp}). Here the average over the momentum spectrum has been performed in the range $[0,5]$ GeV. We observe the same ordering that we have highlighted in Fig. \ref{Fig:v2_charm_pWall_pA_pp}: the smaller is the number of participants, or the value of $\mu$, the smaller is the $v_2$ acquired by the charm quarks. Once again, we note that the value of the integrated $v_2$ for pA collisions at 0-5$\%$ (purple) is compatible with the p-wall result at $\mu=0.5$ GeV (yellow).

\begin{figure}[ht]
\centering
\includegraphics[width=\linewidth]{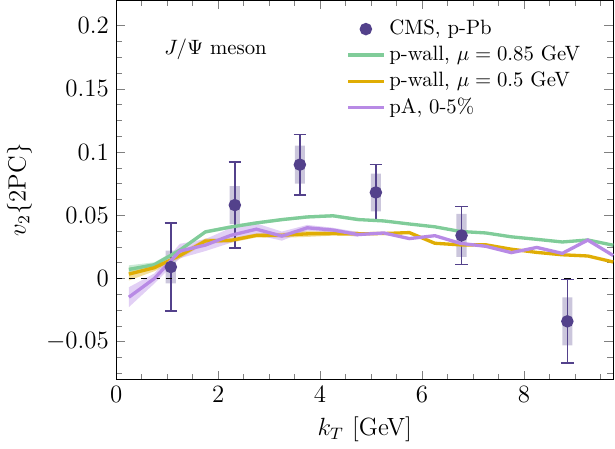}
\caption{
Experimental data of $v_2$ for the $J/\psi$, obtained in p-Pb collisions at $\sqrt{s}_{NN}=8.16$ by the
CMS collaboration at LHC~\cite{CMS:2018duw}. In the experimental data the error bars correspond to statistical uncertainties, while the shaded areas denote the systematic uncertainties. We compare these data with our glasma calculations, computed at $\tau=0.4$ fm/c for several collision systems. In order to make this comparison, we have assumed a naive coalescence picture for the charm quarks \cite{Kolb:2004gi},
which implies that $v_2(k_T)_{J/\psi} = 2 v_2(k_T/2)_{c}$.}
\label{Fig:v2_charm_ExpData}
\end{figure}

It is interesting to estimate the impact of the $v_2$ formed in 
the early stage on the final state observable.
To this end,
in Fig.~\ref{Fig:v2_charm_ExpData} we show the $v_2$ obtained 
in our calculations along with the experimental data 
of $J/\psi$ for p-Pb collisions at
$\sqrt{s_{NN}}=8.16$ TeV measured by the
CMS collaboration at LHC~\cite{CMS:2018duw}. 
In order to compare our calculations with the experimental results, we have assumed a naive coalescence picture~\cite{Kolb:2004gi} 
which implies that the $v_2$ of the $J/\psi$ is obtained from the $v_2$ of the $c$ quarks as
\begin{equation}
    v_2^{J/\psi}(k_T) = 2\, v_2^{c}(k_T/2).
    \label{eq:naive_coalescence}
\end{equation}
Our assumption of pure coalescence is justified by the fact that the charm quark is heavy, hence the fragmentation contribution to the hadronization process is small at intermediate $k_T$, due to the high energy required for the process to occur. The fragmentation processes may become important from $k_T\gtrsim$ 6--7 GeV onwards, and their contribution, here not considered, may describe the decreasing trend observed in experimental data. In any case, the results in Fig. \ref{Fig:v2_charm_ExpData} show that a large part of the final state $v_2$ can be reproduced already from the diffusion of the charm quarks in the initial stage. This highlights the significant effect that the initial glasma fields can have on the momentum anisotropy which is built-up during a proton-ion collision.

\section{Conclusions}
We computed the elliptic flow of charm quarks in the early stage of high-energy proton-nucleus collisions at the LHC energy. We modeled the early stage by the evolving glasma picture, in which the initial condition is made of intense, over-occupied coherent gluon fields, that evolve according to the classical Yang-Mills equations. We adopted a gauge-invariant formulation of the equations of motion of the gluon fields on a real-time lattice. As one of the improvements with respect to previous works on this subject, we included nucleonic fluctuations on both the colliding proton and the nucleus. We then coupled the charm quarks to the glasma via relativistic kinetic theory, namely via the Wong equations, that we solve on top of the evolving gluon fields.

By assuming a well-known form for the gluon spectrum in the initial stages, we evaluated the elliptic flow, $v_2$, of the glasma.
The origin of the glasma $v_2$ is not hydrodynamic, but instead intrinsic to the glasma fields themselves, since it is generated by the relative orientation (in momentum space) of color domains of size $Q_s^{-1}$. 
One of our results is that
the integrated $v_2$ of the bulk in the early stage
decreases
with the number of participant nucleons, $N_\mathrm{part}$. 
This is interpreted by noticing that the larger $N_\mathrm{part}$,
the more isotropic is the distribution of the correlation domains
in the transverse plane, resulting on average in a smaller
anisotropy.

We then addressed the main goal of this work, namely the computation of the elliptic flow coefficient $v_2$ of charm quarks produced and propagated in the early stage of proton--nucleus collisions. 
The charm quark $v_2$ was extracted employing both the event-plane and the two-particle correlation methods, yielding consistent results within uncertainties. 
We found that the evolving glasma fields convey a sizeable azimuthal anisotropy to charm quarks, indicating an efficient transfer of early-time momentum-space anisotropy from the gluon sector to the heavy flavor sector. 
We further studied the dependence of $v_2$ on the number of participants $N_\mathrm{part}$ and observed a clear increase of the charm quark $v_2$ with increasing $N_\mathrm{part}$. 
This trend can be naturally understood as a consequence of the stronger (on average) gluon fields generated in events with larger $N_\mathrm{part}$, which enhance the color Lorentz force acting on heavy quarks and therefore lead to a more effective imprinting of the glasma anisotropy onto the charm distribution.

Finally, moving on to the main result of our work i.e. Fig. \ref{Fig:v2_charm_ExpData}, we compared our results with the available experimental measurements of the $J/\psi$ elliptic flow reported by the CMS Collaboration~\cite{CMS:2018duw}. 
To this end, we adopted the simplest coalescence ansatz,
$v_2^{J/\psi}(k_T) = 2\, v_2^{c}(k_T/2)$, 
which allows one to relate the charm quark anisotropy to that of the formed quarkonium state~\cite{Kolb:2004gi}. 
We find that, despite the short lifetime of the pre-equilibrium stage, charm quarks can acquire a sizeable $v_2$ as a consequence of the large energy density stored in the glasma fields. 
Remarkably, the resulting $v_2^{J/\psi}$ is in the same ballpark of the experimental data. 
Within the limitations of this naive coalescence picture and neglecting any subsequent medium effects, our findings suggest that a substantial fraction of the measured $J/\psi$ elliptic flow in pA collisions may originate from the early-time dynamics. An interesting improvement in this sense would consist in analyzing in which measure the $v_2$ reproduced in the initial stages is retained during a subsequent QGP dynamics: this later stage can be, for instance, modeled via a relativistic Boltzmann evolution \cite{Nugara:2024net}. Research in this direction is underway.

\section{Acknowledgements}
M. R. acknowledges 
Francesco Acerbi,
Bruno Barbieri and John Petrucci for inspiration.
This work has been partly funded by the
European Union – Next Generation EU through the
research grant number P2022Z4P4B “SOPHYA - Sustainable Optimised PHYsics Algorithms: fundamental physics to build an advanced society” under the program PRIN 2022 PNRR of the Italian Ministero dell’Università
e Ricerca (MUR), and by PIACERI “Linea di intervento 1” (M@uRHIC) of the University of Catania. V.G. acknowledges the support PRIN2022 (Project code 2022SM5YAS) within Next Generation EU fundings.

\section{Data availability}
The data that support the findings of this article are openly available \cite{dataset_Elliptic_flow}.

\bibliographystyle{apsrev4-2}
\bibliography{biblio}

\end{document}